# Chern-insulator phase in antiferromagnets

Yuntian Liu[1], Jiayu Li[1] and Qihang Liu[1,2,*]

[1]*Department of Physics and Shenzhen Institute for Quantum Science and Engineering (SIQSE), Southern University of Science and Technology, Shenzhen 518055, China*

[2]*Guangdong Provincial Key Laboratory for Computational Science and Material Design, Southern University of Science and Technology, Shenzhen 518055, China*

[*]Email: liuqh@sustech.edu.cn

## Abstract

The long-sought Chern insulators that manifest quantum anomalous Hall effect are typically considered to occur solely in ferromagnets. Here, we theoretically predict the realizability of Chern insulators in antiferromagnets, of which the magnetic sublattices are connected by symmetry operators enforcing zero net moment. Our symmetry analysis provides comprehensive magnetic layer point groups that allow antiferromagnetic (AFM) Chern insulators, and reveals that in-plane magnetic configuration is required. Followed by first-principles calculations, such design principles naturally lead to two categories of material candidates, exemplified by monolayer $RbCr_4S_8$ and bilayer $Mn_3Sn$ with collinear and noncollinear AFM orders, respectively. We further show that the Chern number could be tuned by slight ferromagnetic canting as an effective pivot. Our work elucidates the nature of Chern-insulator phase in AFM systems, paving a new avenue for the search and design of quantum anomalous Hall insulators with the integration of non-dissipative transport and the promising advantages of the AFM order.

*Introduction*

Chern insulators manifest quantum anomalous Hall effect (QAHE) that has a non-dissipative edge state induced by spontaneous magnetization and non-trivial band topology, hosting the potential for ultra-low power electronic devices[1, 2]. At present, QAHE has been realized in several systems with long-range ferromagnetic (FM) order, such as topological insulators with magnetic doping[3], intrinsic magnetic topological materials[4, 5], and orbital ferromagnetism in twist bilayer graphene[6]. However, due to the scarcity of ferromagnetic insulators and their relatively low Curie temperatures[3-5], the ideal material platform to realize QAHE remains a long-sought challenging issue. On the other hand, antiferromagnetic (AFM) insulators have a series of advantages in device applications, including higher transition temperature, insensitivity to disturbing magnetic fields, zero stray field, high switching speed, etc.[7, 8]. Recent progress in AFM spintronics has revealed a lot of exotic phenomena that indeed exist in antiferromagnets but are typically considered to occur solely in ferromagnets, such as spin splitting[9-11], anomalous Hall effect[12-14], Weyl semimetal phase[15, 16]. Therefore, whether QAHE can be realized in antiferromagnets is an intriguing open question, which, if realized, would bring a paradigm shift in the physical understanding and material selection of QAHE, with extra potential advantages of antiferromagnets mentioned above.

It should be clarified that the classical definition of AFM, proposed by Néel[17], is not merely having zero net magnetic moment, but two (or more) sublattices that carry opposite local moments are crystallographically equivalent[17-19]; in other words, the sublattices need to be connected by certain symmetry operators [see Figs. 1(a) and 1(b)]. In contrast, if the magnetic sublattices are not related by any symmetry operator, even the net magnetic moments are zero in some materials, they are still special cases of ferrimagnets, dubbed as compensated ferrimagnets[20, 21]. Up to now, all the attempts to predict Chern insulators in zero-moment magnets are indeed compensated ferrimagnets themselves[22], or destroy the symmetry connecting the AFM sublattices, thus undergoing a phase transition from antiferromagnets to ferrimagnets[23-27]. Unlike ferrimagnets where the Berry curvatures of different sublattices are not necessarily the same, the symmetry that protects the AFM order usually compensates the Berry curvatures of the sublattices with opposite moments simultaneously [e.g., the combination of inversion *P* and time-reversal *T*, see Fig. 1 (a)], leading to a total Chern number of zero. Therefore, the key challenge to realize Chern-insulator phase in antiferromagnets requires a complete survey of symmetries that compensate the magnetic moments yet retain the Berry curvature [see Fig. 1 (b)].

In this paper, by combining symmetry analysis and first-principles calculations, we treat the search of AFM Chern insulators as a design problem and provide generic design principles as well as corresponding material candidates. The symmetry analysis reveals two crucial design principles for AFM Chern insulators, i.e., (1) the magnetic layer point groups (MLPGs) belong to Table 1; (2) in-plane magnetic configuration. For material realization, we consider two distinct categories, i.e., collinear and noncollinear AFM orders, exemplified by monolayer RbCr$_4$S$_8$ and bilayer Mn$_3$Sn, respectively. Moreover, the ferromagnetic canting induced by external magnetic field could be used as a pivot to manipulate the Chern number. Our work verifies the realizability of QAHE in AFM systems with comprehensive searching principles, which paves a new avenue for the material selection of Chern insulators with promising possibility to integrate the advantages of the AFM order.

*Design Principles*

To illustrate the symmetry conditions of non-zero Chern number, we start from the transformation of Berry curvature in 2D BZ under various symmetries. In 2D, Berry curvature is a single-axis vector defined in the direction perpendicular to the 2D plane ($\Omega^z$), and is norm-preserved under any orthogonal transformations. Therefore, under a symmetry operation $g$ that belongs to the magnetic layer point group (MLPG) $G$ of a given material, the transformed Berry curvature will be either reversed [Fig. 1(a)] or unchanged [Fig. 1(b)]. We summarize the symmetry conditions that reverse or maintain Berry curvatures in the following:

$$\Omega^z(\boldsymbol{k}) = \begin{cases} -\Omega^z(g\boldsymbol{k}) & g \in S \\ \Omega^z(g\boldsymbol{k}) & g \notin S \end{cases} \quad (1)$$

where $S = S_0 \cup PS_0$, $S_0 = \{T, C_{2x}, TC_{2z}, TC_{3z}, TC_{4z}, TC_{6z}\}$ (*P*: space inversion; *T*:

time reversal; the subscripts *x* and *z* represent the rotation axis parallel and perpendicular to the 2D plane, respectively). For the first condition in Eq. (1), the Chern number $C = \frac{1}{2\pi}\int_{BZ}\Omega^z(\boldsymbol{k})d\boldsymbol{k}$, where $\Omega^z(\boldsymbol{k}) = \sum_n f_{kn}\Omega_n^z(\boldsymbol{k})$ represents the sum of Berry curvature of all occupied bands[28, 29], must be 0 due to the compensating $\Omega^z(\boldsymbol{k})$ and $\Omega^z(g\boldsymbol{k})$. Therefore, the corresponding materials could manifest quantum spin Hall effect instead of QAHE. On the other hand, for the materials that fulfill the second condition in Eq. (1), non-zero Chern numbers are naturally allowed.

In addition, we note that the target MLPGs possess symmetries that connect different magnetic sublattices to ensure the AFM nature, so MLPGs $1$ and $\bar{1}$ are excluded. In Table 1, we summarize all the MLPGs supporting non-zero Chern numbers in terms of Laue groups[30]. For a given MLPG, we also list the specific operators connecting the two AFM sublattices, and the permitted spin configuration types, including collinear and non-collinear. Since that both of the magnetic moment and Berry curvature are axial vectors and are odd with respect to *T*, the symmetry operators that can reverse the out-of-plane magnetic moments inevitably reverse Berry curvature $\Omega^z$ as well, leading to zero Chern number. Consequently, the spin configuration in AFM Chern insulators must be in-plane, i.e., collinear, or non-collinear but coplanar. In Supplementary Note I, we use the MLPG $\bar{3}m'$ as an example to illustrate the derivation of Table 1.

Based on the above analysis, we provide two leading design principles of AFM Chern insulator: (1) the MLPG belongs to Table 1; (2) in-plane magnetic configuration. In addition, the magnetic atoms at the same Wyckoff position should be connected by the operators listed in Table 1, with their magnetic moments adding up to zero. These principles, valid for materials either recorded in existing databases or artificially designed, can be used as filters to conduct the efficient material design for QAHE. By using first-principles calculations, we next propose two candidate materials, monolayer $RbCr_4S_8$ and bilayer $Mn_3Sn$ with collinear and noncollinear AFM orders, respectively, to illustrate the topological properties of AFM Chern insulators.

*Collinear antiferromagnets*

We choose monolayer RbCr$_4$S$_8$ as a candidate of collinear AFM, as shown in Fig. 2(a). Bulk RbCr$_2$S$_4$ has an orthorhombic crystal structure with Cr$_2$S$_4$ layers connected by Rb atoms stacking along the c axis, with a space group *Pmmn* and lattice constants *a* = 6.004 Å, *b* = 6.947 Å and *c* = 15.973 Å[31]. By breaking Rb-S bonds, we could get a slab structure of bilayer Cr$_2$S$_4$ with one layer of Rb atoms between them, forming a non-stoichiometric monolayer RbCr$_4$S$_8$ as shown in Fig. 2(a). We consider an in-plane A-type AFM magnetic configuration on Cr along the *a* axis, rendering a meta-stable state with the total energy 2.0 meV/f.u. higher than that of the ground state (Supplementary Note II). As a result, the monolayer RbCr$_4$S$_8$ possess a MLPG $m'm2'$ with mirror reflection $m_z$ connecting two magnetic sublattices, which is included in Table 1.

The electronic structure and topological properties of monolayer RbCr$_4$S$_8$ are shown in Figs. 2(b)-(d). Because the two Cr$_2$S$_4$ layers connected by $m_z$ are intercalated by a Rb buffer layer, the hybridization between these two electronic-active layers is rather weak, leading to almost doubly degenerate bands near the Fermi level. We find that there is a non-trivial gap located at 90 meV above the Fermi level [Fig. 2(b)]. In contrast, the band structure has a trivial gap without spin-orbit coupling (SOC) [Fig. S5(b) in Supplementary Material], indicating that the role of SOC is to induce the band inversion and large Berry curvature near the X point. The calculations of the Wilson loop [Fig. S3(a) in Supplementary Material] and the two chiral edge states [Fig. 2(c)] further confirm that the Chern number at this non-trivial gap is $C = 2$. Furthermore, the distribution of Berry curvature in reciprocal space [Fig. 2(d)] shows that the peaks of Berry curvature appear in the vicinity of the X point and are not eliminated by the AFM symmetry, which is consistent with our prediction.

According to our design principles and the characteristics of weak coupling between magnetic sublattices in monolayer RbCr$_4$S$_8$, we propose a scenario of designing collinear AFM Chern insulator based on the stacking of in-plane FM Chern insulators[32-35]. The key point is to build an AFM bilayer system with FM building blocks connecting each other by mirror symmetry $m_z$ (Supplementary Note IV). Specifically, we provide a hexagonal lattice model Hamiltonian with in-plane Zeeman field to describe such a

designed structure:

$$\hat{H}_{mo} = -t\sum_{\langle i,j\rangle} c_i^\dagger c_j + i\lambda_I \sum_{\langle\langle i,j\rangle\rangle} c_i^\dagger[(\hat{\boldsymbol{E}}_{ij}\times\hat{\boldsymbol{d}}_{ij})\cdot\boldsymbol{\sigma}]c_j$$

$$-i\lambda_R \sum_{\langle\langle i,j\rangle\rangle} c_i^\dagger[(\hat{\boldsymbol{E}}_{ij}\times\hat{\boldsymbol{d}}_{ij})\cdot\boldsymbol{\sigma}]c_j + t_M \sum_i c_i^\dagger(\hat{\boldsymbol{m}}_i\cdot\boldsymbol{\sigma})c_i \quad (2)$$

$$\hat{H}_{bi} = \begin{pmatrix} \hat{H}_{mo} & \hat{H}_i \\ \hat{H}_i^\dagger & \hat{P}(m_z)\hat{H}_{mo}\hat{P}(m_z)^{-1} \end{pmatrix} \quad (3)$$

where $\hat{H}_{mo}$ is the Hamiltonian of a single FM layer with nearest-neighbor hopping, next nearest-neighbor SOC, and on-site in-plane magnetization terms; $\hat{H}_{bi}$ represents the stacking of the AFM bilayer through $m_z$, with $\hat{H}_i$ the coupling between the two FM building blocks and $\hat{P}(m_z)$ the representation of $m_z$. The band structure of such an AFM bilayer system is shown in Fig. S7 in Supplementary Material. We find that when $\hat{H}_i \approx 0$, AFM bilayer is almost two copies of in-plane FM Chern insulator, similar to the case of monolayer RbCr$_4$S$_8$. The mirror symmetry ensures that the total Chern number is the superposition rather than compensation of that of the building blocks. For finite $\hat{H}_i$, as long as the interlayer interaction is insufficient to close the gap, the non-trivial Chern number remains. There are indeed many ways to manipulate various interlayer interactions in realistic materials[36-39]. In addition to some Van der Waals layer materials with intrinsic weak interlayer coupling, it can also be realized by intercalating some inert medium (e.g. BN sheets) into the AFM bilayer, providing an effective method to design collinear AFM Chern insulators from known in-plane FM Chern insulators.

*Non-collinear antiferromagnets*

We choose bilayer Mn$_3$Sn as a candidate material for non-collinear AFM Chern insulator. Bulk Mn$_3$Sn has a hexagonal crystal structure with space group $P6_3/mmc$ and lattice constants $a = b = 5.591$ Å and $c = 4.503$ Å[31]. It is widely studied due to the coexistence of AFM order and the anomalous Hall effect and Weyl semimetal phase, etc.[16, 40-42]. We adopt bilayer Mn$_3$Sn with [0001] surface plane and the all-in-all-out magnetic configuration shown in Fig. 3(a), which is only 4.5 meV/f.u. higher than the energy of the magnetic ground state (Supplementary Note II). The corresponding MLPG is $\bar{3}m'$ with six magnetic Mn atoms in one unit cell connected by $PC_{3z}$ symmetry, which conforms to our design principles in Table 1.

As shown in Fig. 3(b), the band structure with atomic orbital projection shows a 47 meV global gap at 0.5 eV below the Fermi level and the band anticrossing between Mn-$d_{xy}$ and other $d$ orbitals near the M point. In comparison, the corresponding band structure without SOC shows a pair of doubly degenerate Weyl points near each M point [Fig. S6(b) in Supplementary Material], indicating that the role of SOC is to gap the Weyl points, forming the source of large Berry curvature. Furthermore, the edge states shown in Fig. 3(c) have three chiral modes connecting the upper and lower slab bands, giving rise to a non-trivial Chern number $C = 3$. The distribution of Berry curvature further illustrates the influence of symmetry. As shown in Fig. 3(d), a pair of Berry curvature peaks with the same sign appear near each M point along the M-K path, originating from the gapped Weyl points connected by $PC_{3z}$ symmetry.

*Ferromagnetic canting*

Because any symmetry operators that can reverse the out-of-plane magnetic moments will inevitably reverse Berry curvature $\Omega^z$ as well, all the MLPGs that allow AFM Chern insulators will not be broken by a FM canting along the $z$ direction. Given that most of Chern insulators in the previous studies are induced by out-of-plane FM moment, the FM canting could greatly affect the topological properties of AFM Chern insulators, thus providing a novel way to regulate the topological phases.

We still choose the bilayer Mn$_3$Sn to elucidate the effect of FM canting. By adjusting the canting angle to control the out-ot-plane FM component, we find that the band gap and the topological nature can be effectively tuned by FM canting, as shown in Fig. 4(a). FM canting along both positive and negative $z$ direction renders a higher total energy, indicating a stable in-plane AFM state. Remarkably, while +$z$ canting increases the non-trivial band gap, -$z$ canting closes the band gap at a canting angle of -4.5° (1.4 meV/f.u. higher than the in-plane AFM state), and then reopens the gap, indicating the occurrence of topological phase transition. We calculated the band structure and the edge states of bilayer Mn₃Sn with -4.5° canting and -8.9° FM canting, corresponding to the critical phase of gap closing and the phase within the gap reopening region, respectively [Fig. 4(b)-(e)]. A band crossing is clearly shown for -4.5° canting [Fig. 4(b)], with the edge states resembling those of a Weyl semi-metal. In comparison, bilayer Mn₃Sn with -8.9° canting manifests three chiral edge states propagating opposite to those of the AFM phase, indicating a Chern number $C = -3$ [Fig. 4(d)-(e)].

The above results indicate that the Chern number of AFM Chern insulator is highly adjustable via out-of-plane FM canting. In the example of bilayer Mn₃Sn, the Chern number can be changed from 3 to -3 by a canting angle of only 4.5°, for which the required external magnetic field or magnetic proximity is much smaller than that required to reverse the magnetic moments in FM Chern insulators. Therefore, the topological phase transition in AFM Chern insulators could be achieved by FM canting with great ease, providing a new way to regulate the quantum anomalous Hall conductivity. Meanwhile, if the required FM canting is small enough, several advantages of AFM would still be maintained.

*Discussion*

Finally, we discuss the differences of physical mechanism of Chern-insulator phase between antiferromagnetism and ferrimagnetism. For ferrimagnets, even with compensated magnetic moments, the absence of symmetry connecting different magnetic sublattices [Fig. 1(c)] naturally leads to inequivalent energy bands. Therefore, the non-trivial Chern number of ferrimagnetic insulator is typically contributed by single magnetic sublattice[22, 27], resembling the origin of QAHE in ferromagnets. In sharp contrast, with the restrictions described in Fig. 1(b) and Table 1, different magnetic sublattices in AFM are connected by symmetry and thus contribute equally to the total Berry curvature (Fig. S5 and S6 in Supplementary Material). Consequently, AFM Chern insulators usually carry large Chern numbers, as exemplified by the two material candidates shown above ($C = 2$ and $C = 3$ for collinear and non-collinear cases, respectively), providing an ideal platform for realizing the non-dissipative transport with multi-channel and large conductance comparing with FM or ferrimagnetic systems. While our material candidates are on the level of proof-of-concept, our design principles could trigger the fast discovery of ideal AFM Chern insulator materials with the further improvement of the 2D magnetic material databases. In addition, we notice that the recent discovered altermagnetism[11, 19], in which the staggered magnetic order leads to zero net magnetization, possesses spin-splitting electronic structure and corresponding $T$-breaking responses. In this sense, both altermagnetism and AFM Chern insulator are explorations of the field of AFM spintronics with emergent effects that are previously thought to exist exclusively in ferromagnets.

In summary, we elucidate the realizability of Chern insulators in AFM systems with specific symmetry connecting the magnetic sublattices, provide comprehensive design principles for efficient material screening, and predict two example materials with collinear and non-collinear AFM order. Comparing with the traditional FM Chern insulators, the AFM counterpart shares various advantages including large Chern number with enhanced Berry curvature from different magnetic sublattices, great tunability by small FM canting, and high switching speed, etc. Our research paves a new avenue for the field of QAHE by involving a large AFM pool that is previously overlooked, which is promising to find the long-sought high-temperature Chern insulators within the continuously enhanced 2D magnetic material database.

## Methods

**First-Principles Calculations.** Our first-principles calculations were carried out by the Vienna ab initio simulation package (VASP)[43] based on the projector augmented wave (PAW) method[44] within the framework of density-functional theory[45, 46]. The exchange-correlation functional was described by the generalized gradient approximation with the Perdew-Burke-Ernzerhof formalism (PBE)[47, 48] with on-site Coulomb interaction Hubbard U = 5 eV for electrons on d-orbital of Mn, respectively. The total energy convergence criteria was set to $1.0\times10^{-6}$ eV and the plane-wave cutoff energy was set to 350 eV. The whole Brillouin-zone was sampled by a 7×7×1 Monkhorst-Pack grid for monolayer $RbCr_4S_8$ and 9×9×1 for bilayer $Mn_3Sn$. The topological edge states and Berry curvature were obtained from a tight-binding Hamiltonian base on the Wannier functions[49, 50] of Cr-d, S-p orbitals for monolayer $RbCr_4S_8$ and Sn-p, Mn-s,d orbitals for bilayer $Mn_3Sn$, and by the iterative Green's function as implemented in WannierTools package [51].

The Berry curvature and Chern number are calculated by Kubo formula[28, 29]:

$$\Omega_n^z(\boldsymbol{k}) = \sum_{n' \neq n} \frac{2Im[\langle kn|\hat{v}_x|kn'\rangle\langle kn'|\hat{v}_y|kn\rangle]}{(\epsilon_{kn}-\epsilon_{kn'})^2}$$

$$\Omega^z(\boldsymbol{k}) = \sum_n f_{kn} \Omega_n^z(\boldsymbol{k})$$

$$C = \frac{1}{2\pi} \int_{BZ} \Omega^z(\boldsymbol{k}) d\boldsymbol{k} \qquad (4)$$

where $|kn\rangle$ and $\epsilon_{kn}$ are the eigenstate and eigenvalue for band $n$ with wave vector $\boldsymbol{k}$; $\hat{v}_i$ ($i = x, y$) is velocity operator along the $i$ direction. On the other hand, the Berry curvatures contributed by different magnetic sublattices can be quantified by "local Berry curvature"[52-54], expressed as:

$$\Omega_{n,A}^z(\boldsymbol{k}) = \sum_{n' \neq n} 2Im \frac{\langle kn|\hat{v}_x|kn'\rangle\langle kn'|\hat{v}_y|kn\rangle \rho_{k,n,n'}^A}{(\epsilon_{kn}-\epsilon_{kn'})^2}$$

$$\Omega_A^z(\boldsymbol{k}) = \sum_n f_{kn} \Omega_{n,A}^z(\boldsymbol{k}) \qquad (5)$$

where $\rho^A$ is the projection matrix onto the magnetic sublattice A. When the symmetry operation $g$ connecting different sublattices in AFM does not reverse the Berry curvature, different sublattices would contribute the same Berry curvature and thus a nonzero Chern number in total.

## Acknowledgements

We thank Prof. Junxue Li for helpful discussions. This work was supported by National Key R&D Program of China under Grant No. 2020YFA0308900, Guangdong Provincial Key Laboratory for Computational Science and Material Design under Grant No. 2019B030301001, and Center for Computational Science and Engineering of Southern University of Science and Technology.

**Table 1.** Magnetic layer point groups (MLPGs) that allow non-zero Chern number and protect AFM, the corresponding symmetry operations ensuring AFM, and the allowed AFM types. The MLPGs are classified in terms of Laue group having inversion symmetry $P$. To avoid confusion, we use international symbols to denote MLPGs, while use Schoenflies symbols to denote symmetry operations.

| Magnetic layer point groups | Operations ensuring AFM | Allowed AFM types |
|---|---|---|
| $2, m, 2/m$ | $C_{2z}, m_z$ | Collinear |
| $2', m', 2'/m'$ | $TC_{2x}, Tm_x$ | Collinear |
| $2'2'2, m'm2', m'm'2, m'm'm$ | $C_{2z}, m_z, TC_{2x}, Tm_x$ | Collinear / Non-collinear |
| $3, \bar{3}$ | $C_{3z}, PC_{3z}$ | Non-collinear |
| $4, \bar{4}, 4/m$ | $C_{4z}, PC_{4z}$ | Non-collinear |
| $6, \bar{6}, 6/m$ | $C_{6z}, PC_{6z}$ | Non-collinear |
| $32', 3m', \bar{3}m'$ | $C_{3z}, PC_{3z}$ | Non-collinear |
| $42'2', 4m'm', \bar{4}2'm', 4/mm'm'$ | $C_{4z}, PC_{4z}$ | Non-collinear |
| $62'2', 6m'm', \bar{6}m'2', 6/mm'm'$ | $C_{6z}, PC_{6z}$ | Non-collinear |

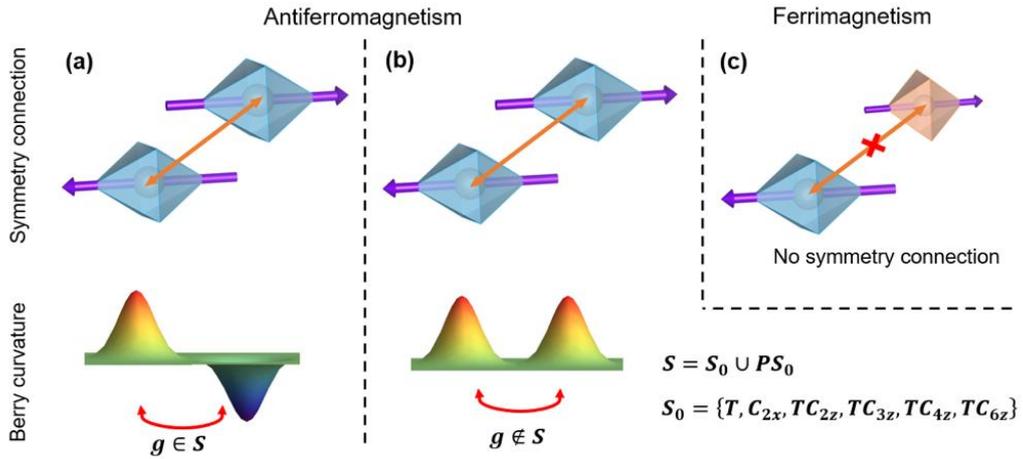

**Fig. 1.** (a-b) Schematic plot of two types of 2D antiferromagnets, i.e., (a) existing symmetry $g \in S$ that reverses Berry curvature and (b) otherwise maintaining Berry curvature ($g \notin S$ for all $g \in G$). The symmetry groups $G$ fulfilling the latter condition allows AFM Chern insulators. Note that both types of AFM need to have symmetry connecting different magnetic sublattices. In contrast, ferrimagnets, even with compensated magnetic moment, do not have equivalent magnetic sublattices connected by symmetry, as shown in panel (c).

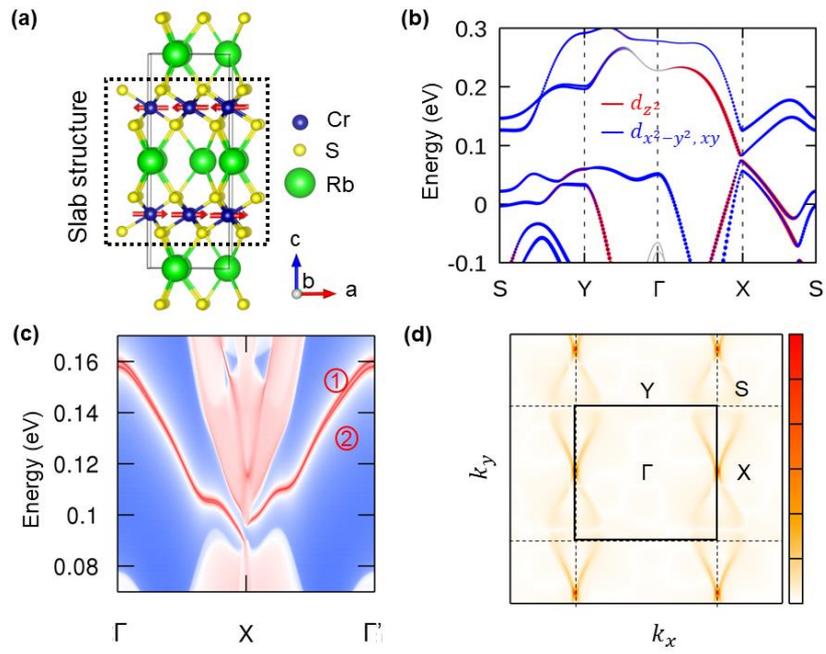

**Fig. 2.** (a) The bulk crystal structure of monolayer $RbCr_4S_8$ and its slab version (dashed box) with the Néel vector along the *a* axis. (b) The slab band structure and (c) edge states of monolayer $RbCr_4S_8$. (d) The distribution of Berry curvature in the reciprocal space, with black lines marking the 2D Brillouin zone.

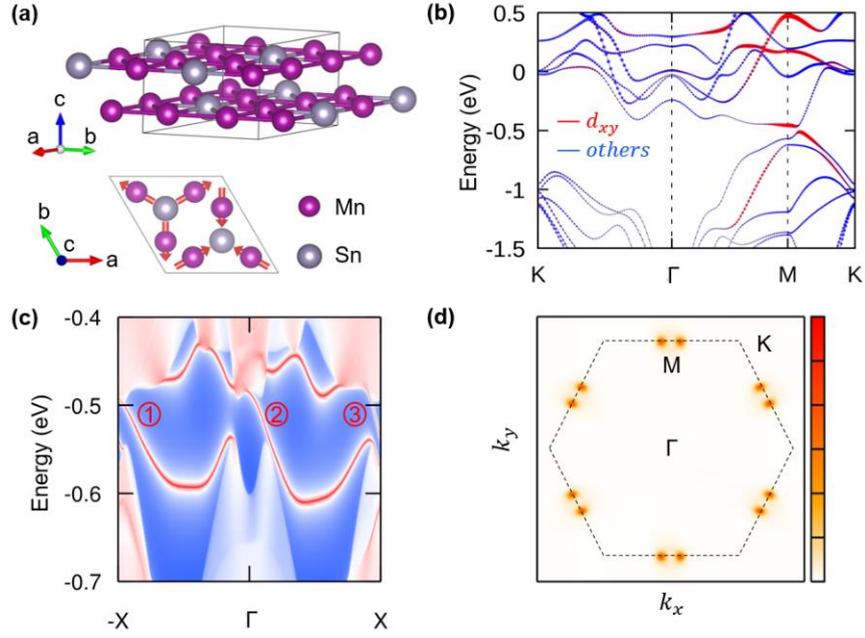

**Fig. 3.** (a) Crystal structure and non-collinear magnetic configuration of AFM bilayer $Mn_3Sn$. (b) The band structure and (c) edge states of bilayer $Mn_3Sn$. (d) The distribution of Berry curvature in reciprocal space, with the dashed lines marking the 2D Brillouin zone.

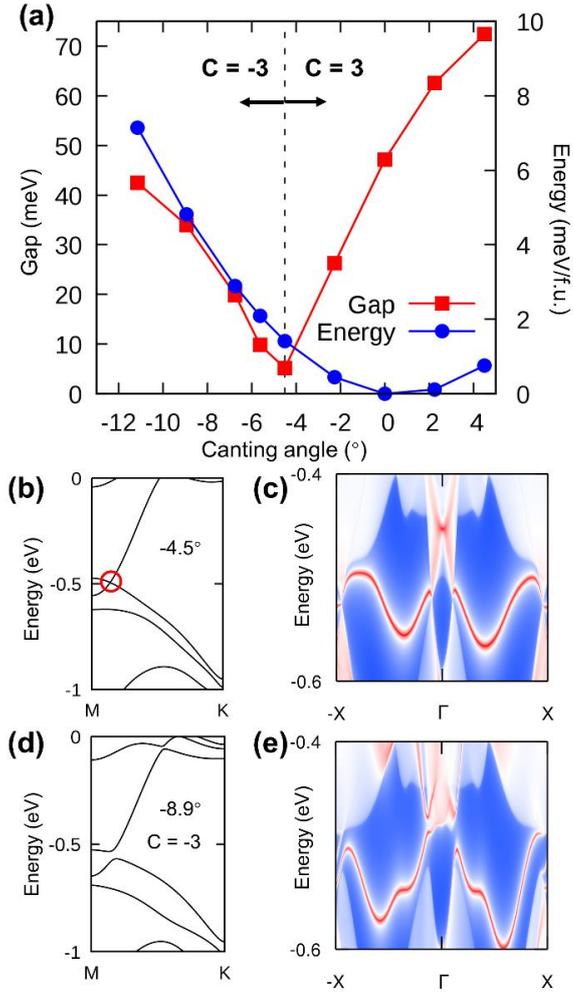

**Fig. 4.** (a) The total energy and the band gap along the M-K path of bilayer $Mn_3Sn$ as a function of the FM canting angle, giving rise to two Chern-insulator phases with $C = 3$ and $C = -3$. (b) The band structure along the M-K path and (c) the edge states with -4.5° FM canting. In panel (b), the band closing point is marked by the red circle. (d-e) Same as (b-c) but with -8.9° FM canting.